# Gender and Research Publishing in India: Uniformly high inequality?[1]

Mike Thelwall, Carol Bailey, Meiko Makita, Pardeep Sud
University of Wolverhampton, Wulfruna Street, Wolverhampton WV1 1LY, UK.
Devika P. Madalli
Documentation Research and Training Center (DRTC), Indian Statistical Institute (ISI), Bangalore 560 059, India.

Women's access to academic careers has been historically limited by discrimination and cultural constraints. Comprehensive information about gender inequality within disciplines is needed to understand the problem and target remedial action. India is the fifth largest research producer but has a low international index of gender inequality and so is an important case. This study assesses gender inequalities in Indian journal article publishing in 2017 for 186 research fields. It also seeks overall gender differences in interests across academia by comparing the terms used in 27,710 articles with an Indian male or female first author. The data show that there are at least 1.5 male first authors per female first author in each of 26 broad fields and 2.8 male first authors per female first author overall. Compared to the USA, India has a much lower share of female first authors but smaller variations in gender differences between broad fields. Dentistry, Economics and Maths are all more female in India, but Veterinary is much less female than in the USA. There is a tendency for males to research thing-oriented topics and for females to research helping people and some life science topics. More initiatives to promote gender equality in science are needed to address the overall imbalance, but care should be taken to avoid creating the larger between-field gender differences found in the USA.
**Keywords**: Research publishing; gender inequality; India; Disciplines; Academic fields

## 1 Introduction

Gender inequalities have been a persistent feature of all modern societies. Although employment-related gender discrimination in various forms is legally prohibited, prejudice and violence against females have not been eradicated. Moreover, gendered social expectations can constrain the career choices of both males and females. Within academia, continuing gender imbalances have been found in many countries (Larivière, Ni, Gingras, Cronin, & Sugimoto, 2013), and particularly at senior levels (e.g., Ucal, O'Neil, & Toktas, 2015; Weisshaar, 2017; Winchester & Browning, 2015). India was the fifth largest research producer in 2017, according to Scopus, but has the highest United Nations Development Programme (UNDP) gender inequality index of the 30 largest research producers in Scopus (/hdr.undp.org/en/data) and so is an important case for global science. Moreover, the complex web of influences that have led to women being underrepresented in science in India is not well understood (Gupta, 2015). The absence of basic information about gender inequalities is a serious limitation because gender issues in India differ from the better researched case of the USA, due to economic conditions, probably stronger family influences (Vindhya, 2007), greater female safety concerns (Vindhya, 2007), and differing cultural expectations (Chandrakar, 2014).

In the USA, where gender imbalances have been much more investigated than in India, underrepresentation of females in science, technology, engineering and mathematics (STEM) fields has caused concern, and is paralleled by underrepresentation of males in health care, elementary education and the domestic sphere (HEED) fields. Many strategies have been suggested to encourage females into STEM subjects. For example, talking about biases faced by women may encourage girls to choose a scientific career (Pietri, Johnson, Ozgumus, & Young, in press) and proactive recruitment and retention strategies at the university and department level may be needed to overcome explicit and hidden biases (Etzkowitz, Kemelgor, & Uzzi, 2000). Current research suggests that these may not address the root causes, however.

The current cause of female STEM underrepresentation in the USA unlikely to be any differences in abilities or direct prejudice (Ceci, & Williams, 2011) but is more likely to be greater male interest in inanimate objects, "things", compared to greater female interest in people (Su, Rounds, & Armstrong, 2009; Su, & Rounds, 2015). Moreover, certain thing-oriented fields have cultures in the USA that are unattractive to females as a career choice, including computer science and engineering (Cheryan, Ziegler, Montoya, & Jiang, 2017). Gender differences in the extent to which people have personal status or social impact goals for their career also explain some gender differences in academic-related career choices (Diekman, Brown, Johnston, & Clark, 2010; Diekman, Steinberg, Brown, Belanger, & Clark, 2017; Diekman, & Steinberg, 2013), overlapping to a large extent with the people/thing theory (Yang & Barth, 2015). International variations in gender differences show that gender influences are not universal. These include the female domination of computing in Malaysia (Othman & Latih, 2006). There are also topics and research methods that are gendered across fields to some extent, including people–oriented methods (female) and abstract methods (male) (Thelwall, Bailey, Tobin, & Bradshaw, submitted).

Despite extensive previous scientometric research into gender differences in academia (Dehdarirad, Villarroya, & Barrios, 2014; Van Den Besselaar, & Sandström, 2016), often focusing on productivity or collaboration within a single area of research (e.g., Nielsen, 2017) and sometimes also a single country (e.g., Araújo & Fontainha, 2017; Frandsen, Jacobsen, Wallin, Brixen, & Ousager, 2015), there have been few systematic investigations into gendered participation rates for individual fields and none for India. This paper reports the first large scale analysis of gender in Indian research publishing, investigating all journal articles published in 2017 and indexed in Scopus with a first author affiliation of India. The goal is to assess evidence of gender inequalities in research publishing overall and within individual fields, as a first systematic step towards characterising the problem. The following research questions drive the study.

- RQ1: In which fields is there a male or female publishing imbalance amongst authors in India?
- RQ2: Which topics and methods are gendered in Indian-authored journal articles?
- RQ3: To what extent does the answer to RQs 1 and 2 echo the situation in the USA?

## 2 Literature review

Gender is a social construct and consists of a set of behaviours and expectations. The two most common genders are male and female and there are also non-binary genders, such as that of the hijra, an iconic gender figure (usually male-bodied self-identified female) in South Asian culture. Hijra have been recently legally recognised as third gender in several South Asian countries, including India (Mal, 2018; Hinchy, 2013). Gender used to be associated with biological sex but these are now regarded as distinct concepts. Behaviours expected from males and females vary between cultures and have changed over time. A hundred years ago, females were not expected to become highly educated, for example.

## 2.1 Gender inequality in India

India is ranked 125$^{th}$ out of 159 countries in the world for gender equality in the United Nations Gender Inequality Index 2015 (UNDP, 2016) and so has a relatively unequal society on a world scale. In contrast, the United States ranks 43$^{rd}$, Switzerland is ranked first (i.e., the most equal society) and Yemen is last (159$^{th}$). The index includes health, empowerment and labour market participation components. According to UN statistics, 27% of Indian females aged 15 or over are employed compared to 79% of males. The corresponding figures for the USA are 56% (female) and 68% (male). From the same source (UNDP, 2016), 35% of females and 61% of males have some secondary education (compared to 95% for both genders in the USA).

India is also ranked below average for the world (108$^{th}$ out of 144) for gender inequality by the World Economic Forum in 2017 (WEF, 2017). India was particularly unequal for labour market participation. In contrast, India is above average (15$^{th}$ out of 144) for female political empowerment because of the relatively high proportion of women in parliament.

Some factors are known to affect the likelihood of females becoming highly educated in India. Girls from wealthier districts are more likely to be educated, although co-residing with in-laws negatively impacts education (Rammohan & Vu, 2018). Girls achieve less at school when they have regular heavy domestic duties at age 12 (Singh & Mukherjee, 2018). Girls from poorer families may be expected to help with housework and childcare, reducing their chance of getting an education (White, Ruther, Kahn, & Dong, 2016). Explicit parental bias towards the education of sons is also an important factor (Singh & Mukherjee, 2018). Girls may sometimes have less desire to be educated (Bhagavatheeswaran, Nair, Stone, Isac, Hiremath, Raghavendra, & Watts, 2016), perhaps because they believe that other factors are more important for their life chances.

## 2.2 Gender in higher education

In contrast with lower levels, the Indian tertiary education gender gap has almost disappeared in recent years. According to Unesco data (Unesco, 2018b; see also: World Bank, 2018), the proportion of females enrolled in tertiary education within India has increased steadily, achieving parity with males for the first time in 2016. Based on government statistics from 2014-15 (the last available), gender inequality is higher for PhDs (41% female) than for undergraduates (47%) (MHRD, 2016, calculations based on data in p.5). Since it takes time for gender equality at lower levels to progress to higher levels, it seems likely that gender parity in PhDs will be achieved in the next five years. Nevertheless, the existing pool of Indians with PhDs is likely to be predominantly male due to their training having taken place during times of higher gender inequality. Moreover, post-PhD gender issues are an important cause of continuing gender inequalities in academic employment, such as a lack of support around childcare commitments (Godbole & Ramaswamy, 2008).

Almost half (49%) of PhDs studied in India in 2016 were in science, technology and engineering, with the social sciences attracting 12% and both management and the Indian language attracting 5% (MHRD, 2016). There is no public information about whether gender differences in PhD program participation vary between fields.

There are substantial differences between India and the USA in the proportion of females studying some subjects. Using Unesco statistics from 2015, 23% of Information and communication technologies tertiary graduates in the USA were female, compared to 46.3% in India (Unesco, 2018a; see also: Nair, 2012). Similarly, for Science, Technology, Engineering and Mathematics (STEM), the female proportion in 2015 was 42% for India and 33% for the USA and for engineering, manufacturing and construction the Indian female proportion in 2015 was 31% and the USA female proportion was 20% (Unesco, 2018). Thus, there seem to be fewer STEM and engineering barriers to females in India than in the USA, or less gender inequality overall in Indian higher education. The root causes include a growing belief in India that computer-related engineering is female-friendly (Varma, 2009), partly because office-based work is relatively safe. Another factor is an increasing parental desire for daughters to have successful careers in the context of plentiful computing jobs in India (Gupta, 2012). Such a career has financial value and many parents believe that it improves marriage prospects (Gupta, 2012; see also: Dutta, 2017). It has been suggested in the past that Indian society is more family focused than is typical for the West, with higher education for women being viewed as a luxury rather than an economic investment because a bride joins the groom's family (Chanana, 2000), but this may no longer be true. Some engineering fields may be regarded internationally as involving an element of dirty, strength-based work, which may be alienating for females (Burke, 2007; Powell, Bagilhole, & Dainty, 2007), but this does not apply to computing.

It is possible that males target engineering and technology careers partly because they offer higher salaries, especially in contrast to the social sciences, arts and humanities (Chanana, 2000).

## 2.3   Gender and research in India

Women are underrepresented amongst higher education faculty in India, particularly at senior levels (Morley & Crossouard, 2015). An investigation of four prestigious higher education institutions with a focus on technology in 2000 found explicit sexist attitudes when hiring staff (Gupta & Sharma, 2003). These included beliefs that women would be less effective due to family commitments.

A study of two Indian Institutes of Technology in 2004 found that women were disadvantaged by male prejudices against them, through being a highly visible minority and having less opportunity for informal interactions because of decorum considerations (Gupta, 2007). The issue of male cultures in Indian technology institutes seems to be decreasing in importance, however (Gupta, 2016).

# 3   Methods

The research design was to download the metadata on all Indian first authored journal articles published in 2017 from Scopus and compare the proportions of male and female first authors in each field. The second stage was to identify words used disproportionately by males or females overall and within individual fields for this article set.

Scopus was chosen as the most comprehensive standard bibliometric database (Falagas, Pitsouni, Malietzis, & Pappas, 2008). In support of this, an approved list of (non-predatory) journals approved by the University Grants Commission for use in promotions and the Academic Performance Indicators (API) in India was taken from Scopus (https://journosdiary.com/2017/01/16/india-ugc-predatory-journals/). Journal articles were analysed because these are the primary outputs in most academic fields and there is not a reasonably comprehensive useful index for conference papers (computing, computational linguistics, some areas of engineering), books (humanities, some social sciences) or artistic works/ performances. This is a limitation of the analysis.

First authors were analysed because these make the largest contribution to a study in most fields (Larivière, Desrochers, Macaluso, Mongeon, Paul-Hus, & Sugimoto, 2016; Yank & Rennie, 1999), even though the first author may tend to be the most senior author in some areas (Kosmulski, 2012). For the word comparisons, terms were analysed in titles, keywords and abstracts since these should summarise the essence of a paper and patterns for these may therefore reveal core gender differences.

Alphabetical ordering occurs in some fields (Araújo & Fontainha, 2017; Levitt & Thelwall, 2013) and may even impact research quality (Joseph, Laband, & Patil, 2005). Nevertheless, it is not universal in any field and is most prevalent in maths, economics, and high energy physics (Waltman, 2012). Alphabetisation may result in the first author of a paper not being its main contributor. This may lead to incorrect gender attributions and reduce apparent gender differences in any analysis. The effect of this is limited by the mixing of alphabetical and non-alphabetical author ordering in any category, the possibility that the main author has the same gender as the first author in an alphabetically ordered article, the possibility that the first author is the main author in an alphabetical ordering, and the occurrence of single-authored articles. A supplementary analysis was conducted to identify the impact of alphabetical ordering. For narrow fields with at least 50 articles, at most 6% (Discrete Mathematics and Combinatorics; Algebra and Number Theory) of articles were estimated to have been assigned an incorrect gender due to alphabetical orderings (see supplementary material online). The net effect on the estimated numbers of male and female first-authored articles is below 1% because male-to-female and female-to-male gender errors tend to cancel out (there are more male first-authored articles but each one is less likely to result in a gender error because most secondary authors are also male). The worst case for a field with at least 50 articles is 2%, for Communication. The gender errors will also tend to slightly reduce the power of the term comparison, but will not lead to spurious positives.

Partly due to alphabetical lists, the corresponding author(s) has also been proposed as a proxy for the main author(s) of an article (Moya-Anegón, Guerrero-Bote, Bornmann, & Moed, 2013; Waltman & van Eck, 2015). This is the person (or persons: Hu, 2009) designated to receive correspondence and may indicate the supervisor, main author, an administrative role (Bhandari, Busse, Kulkarni, Devereaux, Leece, & Guyatt, 2004), the senior author, or an author with a reliable long-term address (Manca, Cugusi, Dvir, & Deriu, 2018). It may also be a junior clerical role if it is expected to lead to routine enquiries or if submitting the manuscript is separate from the research (Leventhal, 2017). There are international and field differences (Liu & Fang, 2014). The corresponding author tends to be the first or last author (Mattsson, Sundberg, & Laget, 2011), however, indicating that it usually connotes main authorship. There is apparently no empirical evidence to justify this statement. Since there is evidence that the first author is usually the main contributor in all broad fields (Larivière et al., 2016), it seems to be reasonable to assume that the first author is the main author even if the corresponding author is not first. An estimated revision of the data based on the assumption that the corresponding author is the main author gives similar gender main author gender shares with the main change to the first table of this paper being that Psychology becomes the most female subject area (see supplementary material). This would not change the main conclusions of the article concerning authorship shares, however. If the main author is more frequently the corresponding author than the first author then the word frequency analysis results would be extended to encompass more terms (because there would be less "noise" in the data due to incorrect gender assignments) but no terms reported in the current paper would be invalid.

Article records were downloaded from Scopus during March 2018 using queries of the form, SUBJMAIN(1108) AND DOCTYPE(ar) AND SRCTYPE(j) AND AFFILCOUNTRY("India"). Here, 1108 is the subject for the Scopus narrow field Horticulture. The query was repeated for all other Scopus narrow field codes, a total of 326, which are grouped into 26 broad fields by Scopus (see the first results table for the broad field names). This query returns articles with any author from India. Results were removed if the first author did not have India as their country affiliation. Thus, the initial dataset was a collection of Scopus journal articles, excluding reviews, from 2017 with a first author affiliation in India.

The gender of each Indian first author was inferred from their first name after scanning Indian baby name websites to extract lists of female and male first names. After discarding 214 names occurring in both lists (e.g., Hari, Kiran), there were 3590 uniquely female and 4180 uniquely male names. These lists were supplemented by a list of US majority male or female names from the US 1990 census when these name genders did not conflict with the Indian names, producing 7383 uniquely female and 5168 uniquely male Indian or American names. The American supplement was added to deal with non-Indian researchers that live in India. The USA is a multi-cultural country so its name list covers many different regions. Authors were assigned a gender when their first name matched a name in either list, otherwise they were left unassigned. Authors listing initials instead of first names were also left unassigned. Journal policies may influence whether authors use initials instead of first names.

The name genders heuristic was checked against a gold standard of author genders. This was created by selecting 1000 articles from the dataset with a random number generator, searching for the home page of the first author (e.g., ResearchGate, Google Scholar or home institution), and checking their gender by inspecting their picture or use of gendered personal pronouns. In 573 (57.3%) cases, a gender was determined. The genders were compared against those produced by the first name list approach described above. For male-authored articles, the automatic method gave a precision of 98.8% and recall of 37.3%, (see Supplement A for calculations and explanations) whereas for the female-authored articles, the automatic method gave a precision of 85.9% and recall of 45.9%. Without the added American names, the precision is slightly higher for females (87.7%) and unchanged for males, but the recall is lower for both females (42.9%) and males (35.9%). Thus, the American addition is an improvement overall, although it makes little difference.

Whilst the method is imperfect, it can identify articles that are much more likely to have been written by females and the same for males, which is the requirement for the word frequency comparisons. In total, 63,351 (39.2%) of the 161,541 articles were allocated a first author gender, or 27,710 (39.7%) of the 69,755 articles after excluding duplicates (articles classified in multiple fields). Many (29%) of the remaining articles had initials or abbreviations rather than first names, from which a gender cannot be detected. Unisex names (e.g. Jyoti) also occurred.

To estimate gender participation proportions for each field, a correction factor was applied to compensate for the inaccuracies in gender detection, which was the precision divided by the recall figure for authors with a gender found online (**female: 1.873; male: 2.651**). Multiplying by precision (the proportion of authors for which the gender was correctly assigned) gives an estimate of the number of authors with a correctly assigned gender. Dividing by recall (the proportion of authors with a correctly identified gender out of all authors with that gender) corrects for share of authors with genders that were not detected from their names. For example, suppose that the precision for females was 90% and the recall was 50% and that the first name mechanism estimated that there were 100 female authors in a field. Multiplying by 90% gives an estimated 90 out of the 100 authors that were correctly identified as female. Dividing by 50% (i.e., 0.5) gives 180 estimated female authors since only half of all female authors are correctly detected by name. This procedure accounts for authors using an ungendered name or initials because they are included in the recall calculation (the manual check of author genders included those with initials instead of first names; their genders could be discovered online from their home page using other information, such as affiliation or article title listed in their CV).

The above procedure gives the estimated number of authors of each gender, correcting for differing accuracies for male and female detection from first names. The greater correction factor for males mainly reflects the lower proportion of males that could be detected by first name checks (i.e., lower recall). The results after multiplying by the correction factor take into account both differing accuracies of gender detection from first names and different rates of including first names (because author genders were identified online for all authors, including those with initials instead of names which may be predominantly female: West, Jacquet, King, Correll, & Bergstrom, 2013). In this sense the corrected figures are unbiased estimates of the correct numbers of male and female first-authored papers, unless there is a gender bias in the proportions of males and females with genders that can be discovered online, which is possible.

The proportion of female authors in each Scopus narrow or broad field is dependent on the Scopus categorisation scheme and may hide more general and methods-based trends. To address this deficiency (RQ2) all terms in the keywords, abstract or titles of papers were analysed to see if they tended to occur more often in male or female first authored texts, using a 2x2 chisquared test for each term (number of documents containing/not containing term vs. male/female first author). A Benjamini-Hochberg correction (Benjamini & Hochberg, 1995) was used to control the Type I familywise error rate for the large number of simultaneous chisquared tests, after discarding terms that were too rare to generate statistically significant results. Plural terms were converted to singular but no stopwords were removed for this analysis because these should occur in similar proportions of articles for each gender (e.g., "the" occurred in 99% of articles for both genders).

A check was also made for issues that were common within fields. From the initial list of 326 fields, 285 had at least one Indian first-authored article. For each of the 186 fields with at least 50 gendered authors, the 20 terms with the high chi squared gender value (as above) were extracted and merged (i.e., generating 186 lists). Terms occurring most frequently were reported as additional potential gender factors.

# 4 Results

The results are compared with the USA (RQ3) primarily in the Discussion.

## 4.1 Research question 1: Field participation

In all 26 broad fields, most journal articles from 2017 with a first author from India were male first-authored (Table 1) and there were almost three male first authors per female first author overall (F/M: 0.35; M/F: 2.83). The fields varied from 0.64 (Nursing) to 0.15 (Veterinary) female authors per male author. For narrow fields with at least 50 Indian first-authored articles in 2017, the proportion varied from 0.05 (Orthopedics and Sports Medicine) to 1.23 (Obstetrics and Gynecology) female authors per male author. Thus, the male dominance of all broad fields has exceptions within narrow fields.

The results cannot be fully analysed with the people/things dimensions because the extent to which researchers perceive each specialism to involve people or things is unknown, as are the key stages at which this information would influence career choices (e.g., childhood ambitions, degree choice, postgraduate decision and topic choice, post-education research/industry career decision, mid-career job change decision). For non-academic careers, this has been assessed with large scale questionnaires (e.g., the O*net 60 item interest profile) but this is impractical for the current paper. Thus, simplifying assumptions must be made for initial judgements about the dimensions.

Complicating the issue of judging the orientation of a field by its subject, a field may be perceived as people/thing oriented if it underpins a career that is perceived as people/thing oriented, it has people/thing research objects or its daily practices involve people or things. Thus, a library and information science researcher may study books (things) to support a career in publishing books (things) or working with authors (people) or library visitors (people). S/he may also enjoy studying books (things) for the stories they tell about people or the opportunity to work alone (no people) or collaboratively (people). For the discussion below, academic fields are accepted at face value for their object of study (e.g., nursing = looking after people; engineering = making things) unless other information is available to support a deeper analysis (e.g., librarians support people's need for books). This is necessarily a superficial and weak approach that does not take into account the Indian context but provides a transparent baseline for analysis that can be challenged.

To increase transparency, three graduate or postdoctoral librarians not involved with this article, two from Asia but not India, were asked to estimate the degree of people/thing orientation of Scopus broad fields and Arts and Humanities (for completeness) and Social Science narrow fields on a five-point scale. Librarians were chosen because this profession records and classifies subject-based information (books, journals, other information resources). Experienced academic librarians from India would have been ideal for this, but could not be found. Agreement by at least two or the average of their results taken as a baseline for the analysis. As the above paragraph suggests, this is a subjective task. Whilst there was strong agreement overall (see Appendix, Table A1 for results and agreement statistics) there was complete agreement on only 20% of the categories and some cases of sharp disagreements, including for Veterinary.

People and health care-based fields (Nursing [1 on the people/thing scale]; Psychology [1]; Dentistry [1]; Arts and Humanities [2]; Health Professions [1]; Social Sciences [2]) have above average levels of female first-authors but Medicine [2] is about average (11 out of 26) and Business, Management and Accounting [2] (16 out of 26) has people-oriented aspects (e.g., personnel management). Thus, the ordering of the broad fields only partly reflects a people/thing dichotomy because some people-related fields do not have a high ranking.

The laboratory-based life sciences broad fields (Immunology and Microbiology [3]; Biochemistry, Genetics and Molecular Biology [4]) also have above average female first author shares despite not being people-dominated, whereas Agricultural and Biological Sciences [4] is 12 out of 26 and Veterinary [2.7] is last. Thus, less laboratory-based life sciences are more male-dominated. Environmental Science [3] (10 out of 26) combines life and physical sciences. Underpinning this, three life-sciences-related narrow fields have the highest proportion of female authors for their broad fields: Bioengineering (Chemistry [5]);

Cellular and Molecular Neuroscience (Neuroscience [3]) and Biomaterials (Materials Science [5]).

At the narrow field level, the female domination of Obstetrics and Gynecology within the Medicine [2] broad field reflects the women's medicine topic, and perhaps female patients' preference for a female doctor. The most female field within Engineering [5] is Architecture, perhaps because its artistic component is perceived as being more feminine or because Architecture is more likely to be office-based and therefore safer. Within Agricultural and Biological Sciences [4], the most female specialism, Food Science, reflects a traditional female role, whereas the most male specialism, Forestry, might be unappealing from a safety aspect for its countryside component. Within Environmental Science [3], the most female-oriented specialism is Waste Management and Disposal. The 180 papers on this theme with a female first author from India frequently deal with microbiology issues, making this a partly life sciences topic. Within Mathematics [5], the most applied specialism, Numerical Analysis, is female-oriented in contrast to the highly abstract most male-oriented Geometry and Topology. This is not due to specific applications since most of these papers were abstract.

Table 1. First author gender ratios for all 26 Scopus broad fields together with the subfields having the highest and lowest ratios of female-authored papers to male-authored papers (for the 186 out of 285 narrow subfields with at least 50 gendered Indian first authored Scopus journal articles in 2017). Broad fields are ranked by F/M ratios. Missing subfields indicate broad fields with fewer than 2 qualifying subfields. F/M odds ratios were multiplied by **1.873/2.651** and M/F odds ratios were multiplied by **2.651/1.873** to correct for first name gender identification biases. USA figures included for comparison (Thelwall, Bailey, Tobin, & Bradshaw, submitted). A complete list is available in the online supplement. P/T: People/things orientation on a scale of 1 to 5, according to three librarians: 1=very people-oriented subject - 5=very thing-oriented subject.

| Broad field | P/T | Sub-fields | India F/M | USA F/M | Most female narrow subfield / Most male narrow subfield | F/M |
|---|---|---|---|---|---|---|
| Nursing | 1 | 16 | 0.64 | 1.93 | Nutrition & Dietetics | 0.60 |
|  |  |  |  |  | - | - |
| Psychology | 1 | 7 | 0.61 | 0.93 | Clinical Psychology | 0.51 |
|  |  |  |  |  | - | - |
| Immunology and Microbiology | 3 | 6 | 0.58 | 0.95 | Microbiology | 0.61 |
|  |  |  |  |  | Immunology & Microbiology (misc) | 0.18 |
| Dentistry | 1 | 5 | 0.57 | 0.33 | Periodontics | 0.75 |
|  |  |  |  |  | Dentistry (misc) | 0.14 |
| Arts and Humanities | 2 | 12 | 0.55 | 0.64 | - | - |
|  |  |  |  |  | - | - |
| Health Professions | 1 | 13 | 0.52 | 0.99 | Radiological & Ultrasound Technology | 0.62 |
|  |  |  |  |  | - | - |
| Biochemistry, Genetics and Molecular Biology | 4 | 15 | 0.51 | 0.67 | Developmental Biology | 0.80 |
|  |  |  |  |  | Endocrinology | 0.32 |
| Social Sciences | 2 | 22 | 0.45 | 0.76 | Health (social science) | 0.92 |
|  |  |  |  |  | Law | 0.27 |
| Pharma, Toxicology and Pharmaceutics | 2.7 | 4 | 0.45 | 0.69 | Pharmaceutical Science | 0.51 |
|  |  |  |  |  | Toxicology | 0.37 |
| Environmental Science | 3 | 11 | 0.44 | 0.52 | Waste Management & Disposal | 0.57 |
| | | | | | Global & Planetary Change | 0.26 |
| Medicine | 2 | 45 | 0.42 | 0.74 | Obstetrics & Gynecology | 1.23 |
| | | | | | Orthopedics & Sports Medicine | 0.05 |
| Agricultural and Biological Sciences | 4 | 11 | 0.40 | 0.49 | Food Science | 0.56 |
| | | | | | Forestry | 0.20 |
| Economics, Econometrics & Finance | 3 | 3 | 0.39 | 0.28 | Economics & Econometrics | 0.41 |
| | | | | | Finance | 0.29 |
| Chemistry | 5 | 7 | 0.37 | 0.39 | Analytical Chemistry | 0.49 |
| | | | | | Chemistry (misc) | 0.24 |
| Chemical Engineering | 5 | 8 | 0.36 | 0.47 | Bioengineering | 0.52 |
| | | | | | Fluid Flow & Transfer Processes | 0.14 |
| Business, Management and Accounting | 2 | 10 | 0.34 | 0.47 | Business & International Management | 0.43 |
| | | | | | Strategy & Management | 0.29 |
| Neuroscience | 3 | 9 | 0.33 | 0.82 | Cellular & Molecular Neuroscience | 0.46 |
| | | | | | Neurology | 0.26 |
| Materials Science | 5 | 8 | 0.33 | 0.39 | Biomaterials | 0.42 |
| | | | | | Materials Science (misc) | 0.27 |
| Physics and Astronomy | 5 | 10 | 0.30 | 0.24 | Radiation | 0.59 |
| | | | | | Nuclear & High Energy Physics | 0.20 |
| Mathematics | 5 | 13 | 0.30 | 0.22 | Numerical Analysis | 0.61 |
| | | | | | Geometry & Topology | 0.16 |
| Earth and Planetary Sciences | 5 | 13 | 0.27 | 0.35 | Paleontology | 0.58 |
| | | | | | Geochemistry & Petrology | 0.20 |
| Computer Science | 5 | 11 | 0.27 | 0.30 | Information Systems | 0.36 |
| | | | | | Artificial Intelligence | 0.22 |
| Decision Sciences | 4 | 3 | 0.24 | 0.32 | Statistics, Probability & Uncertainty | 0.26 |
| | | | | | Management Sci. & Operations Research | 0.23 |
| Energy | 5 | 4 | 0.23 | 0.26 | Renew. Energy, Sustainability & Environ. | 0.29 |
| | | | | | Energy Engineering & Power Technology | 0.16 |
| Engineering | 5 | 15 | 0.22 | 0.32 | Architecture | 0.43 |
| | | | | | Computational Mechanics | 0.09 |
| Veterinary | 2.7 | 4 | 0.15 | 1.49 | - | - |
| | | | | | - | - |

The social sciences broad field (Table 2) encompasses a wide range of different specialisms, not all of which may be perceived as people-oriented. The most female-oriented narrow field, Health (Social Science) [1], is a caring field and Education [1] (3 out of 10) is a nurturing field; Development [3] (4/10) can be nurturing, economic or political. Four fields based on management or control are the most male-oriented (Transportation [3]; Geography, Planning and Development [3]; Political Science and International Relations [2]; Law [2]). The mid position of Library and Information Sciences [3] (6/10) might be surprising, given that libraries play a nurturing role, although it was perceived by the librarian classifiers as a balanced people/thing field. Some of the articles in this category were mathematical, focusing on bibliometrics/scientometrics (44 out of 258, e.g., Garg & Tripathi, 2017), or data/information theory (36, including 13 in *IEEE Transactions on Information Theory,* 11 in *Journal of Chemical Information and Modeling* and 6 in *International Journal of Data Mining and Bioinformatics*) rather than relating to libraries. The rank 2 field Cultural Studies [2] (2/10) includes articles that mainly deal with culture in an Indian context (23 out of the 30 female first-authored papers contained the terms India or Indian) and is a relatively people-oriented field.

Table 2. Ratio of papers by first author gender for Social Sciences Scopus narrow fields (qualification: at least 50 gendered Indian first authored Scopus journal articles in 2017). F/M odds ratios were multiplied by **1.873/2.651** and M/F odds ratios were multiplied by **2.651/1.873** to correct for first name gender identification biases. USA figures included for comparison (Thelwall, Bailey, Tobin, & Bradshaw, submitted). P/T: People/things orientation on a scale of 1 to 5, according to three librarians: 1=very people-oriented subject - 5=very thing-oriented subject.

| Narrow Field | P/T | India F/M | USA F/M |
|---|---|---|---|
| Health (Social Science) | 1 | 0.92 | 1.45 |
| Cultural Studies | 2 | 0.71 | 0.83 |
| Education | 1 | 0.54 | 1.17 |
| Development | 3 | 0.46 | 0.58 |
| Sociology and Political Science | 2 | 0.44 | 0.65 |
| Library and Information Sciences | 3 | 0.37 | 1.14 |
| Transportation | 3 | 0.34 | 0.54 |
| Geography, Planning and Development | 3 | 0.33 | 0.56 |
| Political Science and International Relations | 2 | 0.33 | 0.31 |
| Law | 2 | 0.27 | 0.53 |

### 4.2 Research question 2a: Gendered topics and methods

Both the cross-field (Table 3, 4) and within field (Table 5, 6) term frequency comparisons identified terms that were disproportionately by one gender in abstract articles, titles or keywords. The gendered terms were organised subjectively into themes to simplify reporting and draw out common factors. Terms were reported on their own as a theme unless they had overlaps in meaning or context, in which case they were clustered together. This was achieved by reading random samples of abstracts of papers containing the terms and lists of words that co-occurred with the selected terms. The most female-oriented theme found was childbirth (Table 3), with female researchers being four times more likely to publish a journal article on this topic than male researchers in India. This represents a specialism of the female-oriented field of Obstetrics and Gynecology, perhaps reflecting a greater interest in this topic by women, because it more directly affects them than men, or patients' preferences for female medical personnel for this activity.

Several of the themes reflect the life sciences (Cancer, Cell biology, Genetics, Antibiotics, Molecular biology, Disease, Plants), some with a medical aspect (Cancer, Antibiotics, Disease). Others reflect people (Women, Gender, Children) or care (Dentistry, Patients) or nurturing (Education).

The wastewater theme (F/M: 3.56) is a surprising inclusion, although it is a partly life science issue (the microbiological aspects) and water treatment is a more important health issue than in the USA (e.g., Kumar, Jain, & Jain, 2014). Similarly, the nanomaterials theme is surprising, given the male domination of Materials Science, and the plants theme and the statistical testing theme ostensibly fit the people/things dichotomy. The surveys theme may reflect females being more likely to investigate a people-related topic, for which surveys are a potential research tool. The statistics theme is often found in health research (e.g., the term *patient* is in 23% of articles containing *significance,* 43% of articles containing *statistically,* and 46% of articles containing *SPSS*). The statistics theme associated to a lesser extent with *questionnaire* (*statistically*: 6%; *significance*: 3%; *SPSS*: 25%).

Table 3. The 93 statistically significant **female**-associated academic terms for the overall Scopus 2017 dataset (after duplicate article elimination) organised by subjectively determined theme. The F:M column reports the percentage of female-authored articles containing the term divided by the percentage of male-authored articles containing the term (not the ratio of female to male authors). The highest value for any term in the list is reported.

| Theme* | F:M | Statistically significant terms** |
|---|---|---|
| **Childbirth** | 4.09 | Infertility, vitro, pregnancy, gestation, uterine, obstetric |
| Wastewater | 3.56 | Effluent, detoxification |
| Dentistry | 3.29 | Dental, periodontal |
| Cancer | 2.98 | Agent, cell, tissue, cancer, tumor, assay, carcinoma, proliferation, immunohistochemistry, smear, stained |
| Cell biology | 2.83 | Molecular, human, serum, cytology |
| **Women** | 2.72 | Women |
| Nanomaterials | 2.45 | Nanomaterials |
| Genetics | 2.45 | Concentration, identified, DNA, marker, gene, sequencing, signaling, mutation |
| Antibiotics | 2.39 | Isolated, screened, bacteria, culture, bacterial, antimicrobial, microbial, isolate, microbe |
| Statistical testing | 2.12 | significance, statistically, SPSS |
| **Education** | 1.97 | Student |
| **Gender** | 1.96 | Gender |
| Molecular biology | 1.79 | Acid, activity, biochemical, binding, protein, enzyme, inflammatory, residue |
| **Surveys** | 1.74 | Questionnaire |
| Disease | 1.70 | Disease, disorder, infection, pathogen |
| Health | 1.65 | Health, healthy |
| **Children** | 1.61 | Children |
| **Food** | 1.51 | Food |
| Patients | 1.45 | Presented, treatment, report, positive, clinical, blood, subject |
| Medical drugs | 1.36 | Drug |
| Plants | 1.27 | Plant |

*Bold themes were also found for the top 100 gendered terms for the USA.
**Terms associated with structured abstracts or otherwise not mainly occurring in a clear academic context are not listed in the table: aim, among, association, background, conclusion, design, e, extract, finally, introduction, level, load, motion, oral, potential, proposed, reserved, right, role, screening.

Many of the male-oriented terms are related to things (metalworking, steel, engines, mechanical properties, machine, black holes) or processes that involve things (heat, fluids, electrical power, manufacturing processes, experiment). There are also some abstract concepts (maths, algorithms). The exception to these patterns is surgery. This is people-oriented and a form of medical care, but is male dominated. There have been allegations of a male-oriented culture in surgery in other countries (Yu, Jain, Chakraborty, Wilson, & Hill, 2012; Webster, Rice, Christian, Seemann, Baxter, Moulton, & Cil, 2016), which is a possible explanation.

Table 4: The 53 statistically significant **male**-associated academic terms for the overall Scopus 2017 dataset organised by subjectively determined theme. Duplicate (beyond 4) and ambiguous terms (used in unrelated contexts) are not shown. The M:F column reports the percentage of male-authored articles containing the term divided by the percentage of female-authored articles containing the term (not the ratio of male to female authors). The highest value for any term in the list is reported.

| Theme* | M:F | Statistically significant terms |
| --- | --- | --- |
| Metalworking | 9.05 | Machining, friction, wear, roughness |
| Steel | 6.04 | Weld, steel, hardness |
| Heat | 4.92 | Exchanger, heat, thermal, temperature |
| Fluids | 4.24 | Reynolds, boundary, flow, fluid, pressure |
| **Engines** | 3.55 | Diesel, cylinder, engine, compression |
| Electrical power | 3.29 | Output, input, power, voltage |
| **Surgery** | 2.76 | Fracture |
| **Maths** | 2.70 | Orthogonal, geometry, finite, numerical, equation, |
| Mechanical properties | 2.57 | Vibration, reinforced, shear, mechanical |
| **Black holes (space)** | 2.19 | Hole |
| **Algorithms** | 2.18 | Solve, algorithm |
| Manufacturing processes | 2.10 | Manufacturing |
| **Speed** | 2.02 | Velocity, speed |
| **Machine** | 1.91 | Operation |
| Experiment | 1.56 | Agreement, experiment, experimental, simulation |

*Bold themes were also found for the top 100 gendered terms for the USA.

### 4.2.1 Research question 2b: Within field general topics and methods

Terms that were female-gendered within multiple fields (Table 5) suggested three microscopic life sciences themes (genetics, molecular biology, and bacteria) and women. In the USA, the only life sciences theme was cell biology and there were many people-related themes, including women.

Table 5: Themes detected by using word association tests to explore the contexts in which **female** gendered terms **within individual fields** were used in titles, abstracts or keywords.

| Theme* | Terms in the top 20 female gendered in at least 10 narrow fields |
| --- | --- |
| Genetics | Identified |
| Molecular biology | Enzyme, binding, gene, mutation, strain |
| Bacteria | Strain |
| **Women** | Women |

*Bold themes were also found for the USA.

Terms that were male-gendered within multiple fields suggested multiple themes (Table 6), mainly echoing the similar results for the USA. In addition to the thing-related and quantitative themes, the human-oriented topic of patients was found in both countries. Although patients may be sometimes treated as abstractions with diseases or problems that are the focus of the studies (e.g., the study "Matched cohort study of external cephalic version in women with previous cesarean delivery" has objective "To evaluate the efficacy and safety of external cephalic version (ECV) among women with previous cesarean delivery."), a similar case could be made for most studies involving people (i.e., the focus is on an issue that involves the people) and so the male patients association is an **exception** to the people rule. The heat and fluid flow themes only occurred for India and are thing-associated (terms like temperature and fever might be human equivalents).

Table 6: Themes detected by using word association tests to explore the contexts in which **male** gendered terms **within individual fields were** used in titles, abstracts or keywords.

| Theme* | Terms in the top 20 male gendered in at least 10 narrow fields |
|---|---|
| Fluid | Flow, pressure, velocity |
| **Patients being treated** | Presented (also partly a general term), who, day |
| Heat | Heat |
| Measurement | Scale, less, average |
| **Abstract theory** | Theory |
| Performance improvement | Improve, optimum |
| **Experiments** | Scenario |

*Bold themes were also found for the USA.

## 5 Discussion

This research has several limitations. Many Indian articles are published in journals not indexed by Scopus and therefore excluded from this study. In the arts, humanities and some social sciences, journal articles are not the primary research outputs, with books, performances or other artefacts being more important. Moreover, scholars in some fields in India may have no imperative to produce tangible research outputs, with their teaching and perhaps consultancy activities being sufficient. The study also assumes that the first author of a journal article does the most work, but this may be less true in India than elsewhere, for example, if senior researchers in India (as previously claimed: Menezes, Sharma, Manipady, & Kanchan, 2006) or international collaborators insist on being the first authors for junior researchers' papers. In 86% of fields with at least 50 articles, there were more female first authors (average: 30%) than subsequent authors (average: 24%), but these figures may conceal some unwarranted male first authorships. Finally, the data from 2017 mixes younger researchers from the current more equal education system with older researchers and the gender differences found may naturally reduce each year as newer researchers join and older researchers retire.

The results are compared to a previous study of the USA using the same methods (Thelwall, Bailey, Tobin, & Bradshaw, submitted) as well as the Indian context.

There are several substantial differences from the USA in the gender composition of broad fields. Whilst in the USA, two broad fields have a female majority (Nursing F/M: 1.93; Veterinary F/M: 1.49), and one has close to gender equality (Health Professions: 0.99), all broad fields have a male majority of over 3 to 2 for India. The difference between broad fields in gender composition is narrower for India than for the USA, however. The broad field with the lowest proportion of females is lower in India (0.15 in India for Veterinary compared to 0.22 in the USA for Mathematics) but there are similar numbers of broad fields with proportions below a third (<0.33) in the USA (7, with additionally Physics and Astronomy: 0.24; Energy: 0.26; Economics, Econometrics and Finance: 0.28; Computer Science: 0.30; Engineering: 0.32; Decision Sciences: 0.32) compared to India (8). Thus, whilst Indian research has a substantial overall male bias, broad research field choice is *less* influenced by gender. This echoes the tertiary education situation in India, as discussed above.

For individual broad fields, there are also substantial differences between the USA and India. Most prominently, the least female broad field in India, Veterinary, is the second most female in the USA. Veterinary science may be less attractive to females in India because there may be safety issues for women working at isolated countryside locations with animals. Similarly, Neuroscience has the fifth highest proportion of female-authored papers in the USA (0.82) but is only seventeenth in India (0.33). A word association analysis of the category for the USA and India found Indian research to focus more on healthcare contexts (e.g., the term *patient* is in 39% of Indian and 23% of US articles) whereas US research focused more on theoretical analyses (e.g., *FMRI* [functional magnetic resonance imaging] and *connectivity* are mentioned more). This does not explain the US-India difference, however, since the Indian topics are more female-associated, despite the lower female ranking.

In the opposite direction, Dentistry has the fourth highest proportion of female-authored research in India (0.57) but is only 19[th] in the USA (0.33). Similarly, Economics, Econometrics and Finance has the 13[th] highest proportion of female-authored research in India (0.39) but is only 23[rd] in the USA (0.28). These two broad fields have higher proportions of female authors in India than the USA, despite the much lower overall proportion of female first authors in India. Word association analyses of the broad fields for the USA and India did not find substantial topic differences between these countries, so the differing rankings seems to reflect internationally differing gender relationships. For example, Dentistry research seems to be similar in India and the USA but it is relatively more attractive to Indian females than to American females. Mathematics is more female in India (0.22) than in the USA (0.30) and is higher ranked for female ratio (20/26 instead of 26/26) and so this is a substantial difference. Mathematics research in India seems to be less theoretical (*prove* is in 6% of Indian and 13% of US articles) and more oriented on assessing existing ideas (*proposed* is in 28% of Indian and 15% of US articles) and so the discipline may differ in practice between the two countries.

The social sciences narrow fields very broadly echo the results from the USA. The biggest exception is that Library and Information Sciences is overall female in the USA but strongly male in India. This may be due to the quantitative element, scientometrics, being relatively more popular in India than the USA. For example, the term *citation* occurs in 4% of US articles and 18% of Indian articles in this category.

There are substantial differences with the USA in terms of gendered topics (i.e., topics covered by a higher proportion of female authors than of male authors). Topics that occupy a higher proportion of female than male first authors for the USA but not India include the following, grouped into related themes. Themes evident in the broad field results are omitted. Recall that F:M is the percentage of female-authored articles containing the term divided by the percentage of male-authored articles containing the term.

- Mothers (USA F:M: 15.02), babies (13.21)
- Interpersonal communication (6.56), talking (6.38)
- Sexual violence (3.94)
- Contraception (3.67)
- Psychiatric illness (5.16), Psychotherapy (3.24)
- Female health (3.53); Nursing (3.26); Health management (2.91), Carers (2.73)
- Interviews (3.34)
- Social inequality (3.06)
- Language community (2.78)

Topics that have a significant female bias for India but not the USA include the following, grouped into related themes.
- Wastewater (India F:M: 3.56)
- Cancer (2.98), Cell biology (2.83), Antibiotics (2.39), Molecular biology (1.79)
- Nanomaterials (2.45)
- Genetics (2.45)
- Statistical testing (2.12)
- Disease (1.70), Health (1.65), Patients (1.45)
- Medical drugs (1.36)
- Plants (1.27)

Topics that have a significant male bias for the USA but not India include the following, grouped into related themes.
- Signal processing (USA M:F: 7.45)
- Scholarly debate (7.32)
- Medical imaging (6.45)
- Economics (5.64)
- Religion (4.20)

Topics that have a significant male bias for India but not the USA include the following, grouped into related themes.
- Metalworking (India M:F: 9.05), Steel (6.04), Mechanical properties (2.57)
- Heat (4.92)
- Fluids (4.24)
- Electrical power (3.29)
- Manufacturing processes (2.10)
- Experiment (1.56)

The country differences could occur because there is less research on these topics in India or the USA (so gender differences are less statistically significant in one country despite similar gender difference between them) or because there is a smaller gender difference in one country than another. With a few exceptions, the female-associated topics for India and the USA are related to people, health, caring or medicine whereas the male-associated topics for both India and the USA are associated with engineering, abstraction or quantitative science and technology. Thus, the country differences may be due to differing levels of importance for the topics between India and the USA, such as for health issues or industrial requirements.

# 6 Conclusions

The results show that research publishing in India is dominated by males in all 26 broad areas of scholarship, at least in terms of first-authored journal articles in Scopus. This may exaggerate the male domination of research publishing in India if there is an element of senior (mainly) male researchers unfairly taking first authorship positions (e.g., Menezes et al., 2006). Research publishing is male-dominated in 183 out of 186 narrow fields, with the exceptions being Histology (F/M: 1.15) (56% for cytology – the study of cells), Reproductive Medicine (F/M: 1.16), and Obstetrics and Gynecology (F/M: 1.23). Indian research has fewer female researchers (at least in terms of Scopus first authors in 2017) than the USA but smaller differences between fields in the share of female researchers. In terms of fields and topics, the data echoes to some extent the people/things dichotomy previously found for the USA (with patients as an important exception), but not the greater female use of qualitative methods found in the USA. Thus, the gender imbalance in India is substantially different from that in the USA but with some overlaps. There are also many differences for individual topics that may be due to national context (e.g., type of commercial activity, differing health issues) that change the nature of a field (e.g., more algorithms in Indian mathematics).

The substantial female minority overall and in all 26 broad areas of scholarship underlines the importance of initiatives to overcome gender inequalities in research within India (e.g., the Indo-U.S. Fellowship for Women in STEMM), although a more systematic effort is still needed (Ovseiko, Godbole, & Latimer, 2017). The differences with the USA confirm that national policies or cultures can influence the gendering of science. This should encourage initiatives to overcome gender imbalances by showing that they are not universal and can therefore be combatted. Indian policy-makers and research managers should also guard against the potential for gender imbalances to increase in areas where the USA has a greater discrepancy in case research cultures from the USA influence practice in India.

# 7 References

Araújo, T., & Fontainha, E. (2017). The specific shapes of gender imbalance in scientific authorships: a network approach. Journal of Informetrics, 11(1), 88-102.

Benjamini, Y., & Hochberg, Y. (1995). Controlling the false discovery rate: a practical and powerful approach to multiple testing. Journal of the Royal Statistical Society. Series B (Methodological), 289-300.

Bhagavatheeswaran, L., Nair, S., Stone, H., Isac, S., Hiremath, T., Raghavendra, T., & Watts, C. (2016). The barriers and enablers to education among scheduled caste and scheduled tribe adolescent girls in northern Karnataka, South India: A qualitative study. International journal of educational development, 49, 262-270.

Bhandari, M., Busse, J. W., Kulkarni, A. V., Devereaux, P. J., Leece, P., & Guyatt, G. H. (2004). Interpreting authorship order and corresponding authorship. Epidemiology, 15(1), 125-126.

Burke, R. J., (2007). Women and minorities in STEM: A primer. In: Burke, R. J., & Mattis, M. C. (Eds.). Women and minorities in science, technology, engineering, and mathematics: Upping the numbers. Cheltenham, UK: Edward Elgar Publishing.

Ceci, S. J., & Williams, W. M. (2011). Understanding current causes of women's underrepresentation in science. Proceedings of the National Academy of Sciences, 108(8), 3157-3162.

...

Chanana, K. (2000). Treading the hallowed halls: Women in higher education in India. Economic and Political Weekly, 35(12), 1012-1022.

Chandrakar, C. L. (2014). Why "Ritiya" could not go to sell vegetables? Myth versus reality in terms of caste, culture and livelihood. Journal of Rural Social Sciences, 29(2), 48-56.

Cheryan, S., Ziegler, S. A., Montoya, A. K., & Jiang, L. (2017). Why are some STEM fields more gender balanced than others? Psychological Bulletin, 143(1), 1-35.

Dehdarirad, T., Villarroya, A., & Barrios, M. (2014). Research trends in gender differences in higher education and science: a co-word analysis. Scientometrics, 101(1), 273-290.

Diekman, A. B., Brown, E. R., Johnston, A. M., & Clark, E. K. (2010). Seeking congruity between goals and roles: A new look at why women opt out of science, technology, engineering, and mathematics careers. Psychological Science, 21(8), 1051-1057.

Diekman, A. B., Steinberg, M., Brown, E. R., Belanger, A. L., & Clark, E. K. (2017). A goal congruity model of role entry, engagement, and exit: Understanding communal goal processes in STEM gender gaps. Personality and Social Psychology Review, 21(2), 142-175.

Diekman, A. B., & Steinberg, M. (2013). Navigating social roles in pursuit of important goals: A communal goal congruity account of STEM pursuits. Social and Personality Psychology Compass, 7(7), 487-501.

Dutta, D. (2017). Cultural barriers and familial resources for negotiation of engineering careers among young women: Relational dialectics theory in an Asian perspective. Journal of Family Communication, 17(4), 338-355.

Etzkowitz, H., Kemelgor, C., & Uzzi, B. (2000). Athena unbound: The advancement of women in science and technology. Cambridge, UK: Cambridge University Press.

Falagas, M. E., Pitsouni, E. I., Malietzis, G. A., & Pappas, G. (2008). Comparison of PubMed, Scopus, web of science, and Google scholar: strengths and weaknesses. The FASEB journal, 22(2), 338-342.

Frandsen, T. F., Jacobsen, R. H., Wallin, J. A., Brixen, K., & Ousager, J. (2015). Gender differences in scientific performance: A bibliometric matching analysis of Danish health sciences Graduates. Journal of Informetrics, 9(4), 1007-1017.

Garg, K. C., & Tripathi, H. K. (2017). Bibliometrics and scientometrics in India: An overview of studies during 1995-2014 Part I: Indian publication output and its citation impact. Annals of Library and Information Studies, 64(1), 28-36.

Godbole, R. & Ramaswamy, R. (2008). Women scientists in India. In: AASSA (ed.), Women in science and technology in Asia. KAST: Gyeonggi-Do, Republic of Korea (pp. 67-84).

Gupta, N., & Sharma, A. K. (2003). Gender inequality in the work environment at institutes of higher learning in science and technology in India. Work, Employment and Society, 17(4), 597-616.

Gupta, N. (2007). Indian women in doctoral education in science and engineering: A study of informal milieu at the reputed Indian institutes of technology. Science, Technology, & Human Values, 32(5), 507-533.

Gupta, N. (2012). Women undergraduates in engineering education in India: A study of growing participation. Gender, Technology and Development, 16(2), 153-176.

Gupta, N. (2015). Gender equality in science in India: an undeveloped agenda. Current Science, 109(1), 9-10.

Gupta, N. (2016). Perceptions of the work environment: The issue of gender in Indian scientific research institutes. Indian Journal of Gender Studies, 23(3), 437-466.

Hinchy, J. (2013). Troubling bodies:'eunuchs,' masculinity and impotence in colonial North India. South Asian History and Culture, 4(2), 196-212.

Hu, X. (2009). Loads of special authorship functions: Linear growth in the percentage of "equal first authors" and corresponding authors. Journal of the American Society for Information Science and Technology, 60(11), 2378-2381.

Joseph, K., Laband, D. N., & Patil, V. (2005). Author order and research quality. Southern Economic Journal, 71(3), 545-555.

Kosmulski, M. (2012). The order in the lists of authors in multi-author papers revisited. Journal of Informetrics, 6(4), 639-644.

Krippendorff, K. (2013), Content Analysis: An Introduction to Its Methodology (3 ed.) Thousand Oaks, CA: Sage.

Kumar, H., Jain, K., & Jain, R. (2014). A study of prevalence of intestinal worm infestation and efficacy of anthelminthic drugs. Medical Journal Armed Forces India, 70(2), 144-148.

Larivière, V., Desrochers, N., Macaluso, B., Mongeon, P., Paul-Hus, A., & Sugimoto, C. R. (2016). Contributorship and division of labor in knowledge production. Social Studies of Science, 46(3), 417-435.

Larivière, V., Ni, C., Gingras, Y., Cronin, B., & Sugimoto, C. R. (2013). Global gender disparities in science. Nature, 504(7479), 211-213.

Leventhal, P. (2017). Can a medical writer submit a manuscript on behalf of a corresponding author?. Medical Writing, 26, 45-46.

Levitt, J. M., & Thelwall, M. (2013). Alphabetization and the skewing of first authorship towards last names early in the alphabet. Journal of Informetrics, 7(3), 575-582.

Liu, X. Z., & Fang, H. (2014). Scientific group leaders' authorship preferences: An empirical investigation. Scientometrics, 98(2), 909-925.

Mal S. (2018). The hijras of India: A marginal community with paradox sexual identity. Indian Journal of Social Psychiatry, 34(1), 79-85.

Manca, A., Cugusi, L., Dvir, Z., & Deriu, F. (2018). Non-corresponding authors in the era of meta-analyses. Journal of clinical epidemiology, 98, 159-161.

Mattsson, P., Sundberg, C. J., & Laget, P. (2011). Is correspondence reflected in the author position? A bibliometric study of the relation between corresponding author and byline position. Scientometrics, 87(1), 99-105.

Menezes, R. G., Sharma, P. S., Manipady, S., & Kanchan, T. (2006). Order of authorship. Current Science, 91(6), 736.

MHRD (2016). Educational Statistics - At a Glance. MHRD. http://mhrd.gov.in/sites/upload_files/mhrd/files/statistics/ESG2016_0.pdf

Morley, L., & Crossouard, B. (2015). Women in higher education leadership in South Asia: Rejection, refusal, reluctance, revisioning. http://sro.sussex.ac.uk/53151/1/morley_crossouard_final_report_22_dec2014.pdf

Moya-Anegón, F., Guerrero-Bote, V. P., Bornmann, L., & Moed, H. F. (2013). The research guarantors of scientific papers and the output counting: a promising new approach. Scientometrics, 97(2), 421-434.

Nair, S. (2012). National Assessment of the Participation of Women and Girls in the National STI System Based on the Gender Equality-Knowledge Society Framework: India report. http://wisat.org/wp-content/uploads/India_GE-KS.pdf

Nielsen, M. W. (2017). Gender and citation impact in management research. Journal of Informetrics, 11(4), 1213-1228.

Othman, M., & Latih, R. (2006). Women in computer science: no shortage here! Communications of the ACM, 49(3), 111-114.

Ovseiko, P. V., Godbole, R. M., & Latimer, J. (2017). Gender equality: Boost prospects for women scientists. Nature, 542(7639), 31.

Pietri, E. S., Johnson, I. R., Ozgumus, E., & Young, A. I. (in press). Maybe she is relatable: increasing women's awareness of gender bias encourages their identification with women scientists. Psychology of Women Quarterly.

Powell, A., Bagilhole, B., & Dainty, A. (2007). The good, the bad and the ugly: women engineering students' experiences of UK higher education. In: Burke, R. J., & Mattis, M. C. (Eds.). Women and minorities in science, technology, engineering and mathematics: Upping the numbers, Cheltenham, UK: Edward Elgar Publishing (pp. 47-70).

Rammohan, A., & Vu, P. (2018). Gender Inequality in Education and Kinship Norms in India. Feminist Economics, 24(1), 142-167.

Singh, R., & Mukherjee, P. (2018). 'Whatever she may study, she can't escape from washing dishes': gender inequity in secondary education–evidence from a longitudinal study in India. Compare: A Journal of Comparative and International Education, 48(2), 262-280.

Su, R., Rounds, J., & Armstrong, P. I. (2009). Men and things, women and people: A meta-analysis of sex differences in interests. Psychological Bulletin, 135(6), 859–884

Su, R., & Rounds, J. (2015). All STEM fields are not created equal: People and things interests explain gender disparities across STEM fields. Frontiers in Psychology, 6, paper 189. doi:10.3389/fpsyg.2015.00189

Thelwall, M., Bailey, C., Tobin, C., & Bradshaw, N. (submitted). Gender differences in research areas and topics: An analysis of publications in 285 fields. https://arxiv.org/abs/1809.01255

Ucal, M., O'Neil, M. L., & Toktas, S. (2015). Gender and the wage gap in Turkish academia. Journal of Balkan and Near Eastern Studies, 17(4), 447-464.

UNDP (2016). Human Development Report 2016: Human Development for Everyone. United Nations. http://hdr.undp.org/sites/default/files/2016_human_development_report.pdf

Unesco (2018a). Percentage of female tertiary graduates by field of education between 1998 and 2016. http://uis.unesco.org/indicator/edu-compl-grad-field_female

Unesco (2018b). India| Unesco UES. http://uis.unesco.org/country/IN

Van Den Besselaar, P., & Sandström, U. (2016). Gender differences in research performance and its impact on careers: a longitudinal case study. Scientometrics, 106(1), 143-162.

Varma, R. (2009). Exposure, training, and environment: Women's participation in computing education in the United states and India. Journal of Women and Minorities in Science and Engineering, 15(3), 205-222.

Vindhya, U. (2007). Quality of women's lives in India: Some findings from two decades of psychological research on gender. Feminism & Psychology, 17(3), 337-356.

Waltman, L., & van Eck, N. J. (2015). Field-normalized citation impact indicators and the choice of an appropriate counting method. Journal of Informetrics, 9(4), 872-894.

Waltman, L. (2012). An empirical analysis of the use of alphabetical authorship in scientific publishing. Journal of Informetrics, 6(4), 700-711.

Webster, F., Rice, K., Christian, J., Seemann, N., Baxter, N., Moulton, C. & Cil, T. (2016). The erasure of gender in academic surgery: a qualitative study. The American Journal of Surgery, 212(4), 559-565.

WEF (2017). The Global Gender Gap Report 2017. World Economic Forum. https://www.weforum.org/reports/the-global-gender-gap-report-2017

Weisshaar, K. (2017). Publish and perish? An assessment of gender gaps in promotion to tenure in academia. Social Forces, 96(2), 529-560.

West, J. D., Jacquet, J., King, M. M., Correll, S. J., & Bergstrom, C. T. (2013). The role of gender in scholarly authorship. PloS ONE, 8(7), e66212.

White, G., Ruther, M., Kahn, J. R., & Dong, D. (2016). Gender inequality amid educational expansion in India: an analysis of gender differences in the attainment of reading and mathematics skills. Journal of Research in Gender Studies, 6(2), 153-182.

Winchester, H. P., & Browning, L. (2015). Gender equality in academia: A critical reflection. Journal of Higher Education Policy and Management, 37(3), 269-281.

World Bank (2018). Education data. http://datatopics.worldbank.org/education/country/india  Accessed 28 May 2018.

Yang, Y., & Barth, J. M. (2015). Gender differences in STEM undergraduates' vocational interests: People–thing orientation and goal affordances. Journal of Vocational Behavior, 91(1), 65-75.

Yank, V., & Rennie, D. (1999). Disclosure of researcher contributions: a study of original research articles in The Lancet. Annals of Internal Medicine, 130(8), 661-670.

Yu, T. C., Jain, A., Chakraborty, M., Wilson, N. C., & Hill, A. G. (2012). Factors influencing intentions of female medical students to pursue a surgical career. Journal of the American College of Surgeons, 215(6), 878-889.

# 8 Appendix

The following instructions were given to the three librarians for the classification task, separately for the broad categories and each set of narrow fields. Text in square brackets was given for either the narrow or the broad field instructions. None of the coders had read any part of the current paper. They may have read the open access USA paper but this seems unlikely since they had not been informed about it. The Arts and Humanities narrow fields and some of the Social Sciences narrow fields are not used in the current paper but are provided for wider usefulness (e.g., similar follow-up papers).

> This task is to provide supporting information for a paper that discusses factors related to the extent to which a research within an academic field involves "people" or "things". Completing this is optional. Your answers will be used in the paper anonymously and combined with the answers of other people. There are no correct answers and please use your personal judgement.
> 
> Please rate (with an X in the appropriate box) the following [broad/narrow] Scopus subjects for the extent to which you believe that they would appeal to a person that prefers to work within "people" or "things" (inanimate/non-living objects). [Please consult the appendix for a list of subfields for each one to help with your decision.]
> 
> 1=very people-oriented subject - 5=very thing-oriented subject.

Table A1. The people/thing categorisation from three graduate or postdoctoral librarians (A, B, C) of all Scopus broad fields (except Multidisciplinary) and all Arts and Humanities and Social Science narrow fields. The final column is used in the paper. It is the category chosen by at least two of the three librarians, otherwise it is the average of all three librarians. Bold categories have a consensus.

| ASJC | Description | A | B | C | Agreed | Average | Final |
|---|---|---|---|---|---|---|---|
| 1100 | Agricultural and Biological Sciences | 4 | 2 | 4 | 4 | 3.3 | 4 |
| 1200 | Arts and Humanities | 3 | 2 | 1 | | 2 | 2 |
| 1300 | Biochemistry, Genetics & Molecular Biology | 4 | 3 | 4 | 4 | 3.7 | 4 |
| 1400 | Business, Management and Accounting | 3 | 1 | 2 | | 2 | 2 |
| 1500 | **Chemical Engineering** | 5 | 5 | 5 | 5 | 5 | 5 |
| 1600 | Chemistry | 5 | 4 | 5 | 5 | 4.7 | 5 |
| 1700 | **Computer Science** | 5 | 5 | 5 | 5 | 5 | 5 |
| 1800 | Decision Sciences | 5 | 4 | 3 | | 4 | 4 |
| 1900 | Earth and Planetary Sciences | 5 | 2 | 4 | | 3.7 | 3.7 |
| 2000 | Economics, Econometrics and Finance | 4 | 3 | 2 | | 3 | 3 |
| 2100 | Energy | 5 | 5 | 4 | 5 | 4.7 | 5 |
| 2200 | **Engineering** | 5 | 5 | 5 | 5 | 5 | 5 |
| 2300 | Environmental Science | 3 | 2 | 3 | 3 | 2.7 | 3 |
| 2400 | Immunology and Microbiology | 4 | 3 | 2 | | 3 | 3 |
| 2500 | **Materials Science** | 5 | 5 | 5 | 5 | 5 | 5 |
| 2600 | **Mathematics** | 5 | 5 | 5 | 5 | 5 | 5 |
| 2700 | Medicine | 3 | 1 | 2 | | 2 | 2 |
| 2800 | Neuroscience | 3 | 1 | 3 | 3 | 2.3 | 3 |
| 2900 | **Nursing** | 1 | 1 | 1 | 1 | 1 | 1 |
| 3000 | Pharmacology, Toxicology & Pharmaceutics | 4 | 1 | 3 | | 2.7 | 2.7 |
| 3100 | Physics and Astronomy | 5 | 4 | 5 | 5 | 4.7 | 5 |
| 3200 | Psychology | 2 | 1 | 1 | 1 | 1.3 | 1 |
| 3300 | Social Sciences | 2 | 2 | 1 | 2 | 1.7 | 2 |
| 3400 | Veterinary | 3 | 1 | 4 | | 2.7 | 2.7 |
| 3500 | **Dentistry** | 1 | 1 | 1 | 1 | 1 | 1 |
| 3600 | **Health Professions** | 1 | 1 | 1 | 1 | 1 | 1 |
| 1202 | History | 2 | 2 | 1 | 2 | 1.7 | 2 |
| 1203 | Language and Linguistics | 2 | 2 | 1 | 2 | 1.7 | 2 |
| 1204 | Archaeology | 3 | 4 | 1 | | 2.7 | 2.7 |
| 1205 | Classics | 2 | 3 | 1 | | 2 | 2 |
| 1206 | Conservation | 4 | 4 | 3 | 4 | 3.7 | 4 |
| 1207 | History and Philosophy of Science | 2 | 4 | 2 | 2 | 2.7 | 2 |
| 1208 | Literature and Literary Theory | 3 | 3 | 1 | 3 | 2.3 | 3 |
| 1209 | Museology | 3 | 3 | 1 | 3 | 2.3 | 3 |
| 1210 | Music | 4 | 1 | 1 | 1 | 2 | 1 |
| 1211 | Philosophy | 2 | 3 | 1 | | 2 | 2 |
| 1212 | Religious studies | 2 | 1 | 1 | 1 | 1.3 | 1 |
| 1213 | Visual Arts and Performing Arts | 2 | 1 | 1 | 1 | 1.3 | 1 |

| Code | Field | A | B | C | D | Avg | F |
|------|-------|---|---|---|---|-----|---|
| 3302 | Archaeology | 3 | 3 | 1 | 3 | 2.3 | 3 |
| 3303 | Development | 3 | 3 | 3 | 3 | 3 | 3 |
| 3304 | **Education** | 1 | 1 | 1 | 1 | 1 | 1 |
| 3305 | Geography, Planning and Development | 4 | 3 | 3 | 3 | 3.3 | 3 |
| 3306 | Health (social science) | 1 | 2 | 1 | 1 | 1.3 | 1 |
| 3307 | Human Factors and Ergonomics | 2 | 1 | 1 | 1 | 1.3 | 1 |
| 3308 | Law | 2 | 2 | 1 | 2 | 1.7 | 2 |
| 3309 | Library and Information Sciences | 3 | 3 | 2 | 3 | 2.7 | 3 |
| 3310 | Linguistics and Language | 2 | 3 | 1 |   | 2 | 2 |
| 3311 | Safety Research | 2 | 2 | 3 | 2 | 2.3 | 2 |
| 3312 | Sociology and Political Science | 2 | 3 | 1 |   | 2 | 2 |
| 3313 | Transportation | 3 | 1 | 3 | 3 | 2.3 | 3 |
| 3314 | Anthropology | 1 | 3 | 1 | 1 | 1.7 | 1 |
| 3315 | Communication | 2 | 1 | 1 | 1 | 1.3 | 1 |
| 3316 | Cultural Studies | 2 | 2 | 1 | 2 | 1.7 | 2 |
| 3317 | Demography | 2 | 1 | 1 | 1 | 1.3 | 1 |
| 3318 | **Gender Studies** | 1 | 1 | 1 | 1 | 1 | 1 |
| 3319 | **Life-span and Life-course Studies** | 1 | 1 | 1 | 1 | 1 | 1 |
| 3320 | Political Science & International Relations | 2 | 2 | 1 | 2 | 1.7 | 2 |
| 3321 | Public Administration | 2 | 1 | 1 | 1 | 1.3 | 1 |
| 3322 | Urban Studies | 2 | 3 | 1 |   | 2 | 2 |
|      | **Average** | **2.9** | **2.4** | **2.1** |   | **2.5** | **2.52** |

Agreement statistics with Krippendorff (2013) multi-coder alpha using distance measure abs(a-b) and Pearson correlations.

- Coders A, B, C: Alpha = 0.534; full agreement: 12/59: 20%.
- Coders: A, B: Alpha = 0.657; pairwise agreement: 25/59 (42%); Pearson: 0.650.
- Coders: A, C: Alpha = 0.683; pairwise agreement: 22/59 (37%); Pearson: 0.835.
- Coders: B, C: Alpha = 0.598; pairwise agreement: 21/59 (36%); Pearson: 0.585.